\documentclass[a4paper]{article}
\usepackage{graphicx,color,amsmath,epstopdf,authblk}
\begin{document}
\title{Practical circuits with Physarum Wires
PREPRINT}

\author[1]{James G. H. Whiting\thanks{Corresponding author: James.Whiting@UWE.ac.uk}}
\author[1]{Richard Mayne}
\author[2]{Nadine Moody}
\author[2]{Ben de Lacy Costello}
\author[1]{Andrew Adamatzky}
\affil[1]{Unconventional Computing Group, University of the West of England, Bristol, UK}
\affil[2]{Institute for Biosensing, University of the West of England, Bristol, UK}


\date{}


\maketitle

\begin{abstract}
\emph{Purpose} 
Protoplasmic tubes of \emph{Physarum polycephalum}, also know as Physarum Wires (PW), have been previously suggested as novel bio-electronic components. Until recently, practical examples of electronic circuits using PWs have been limited. These PWs have been shown to be self repairing, offering significant advantage over traditional electronic components. This article documents work performed to produce practical circuits using PWs.\\
\emph{Method}
We have demonstrated through manufacture and testing of hybrid circuits that PWs can be used to produce a variety of practical electronic circuits. A purality of different applications of PWs have been tested to show the universality of PWs in analogue and digital electronics.\\
\emph{Results}
Voltage dividers can be produced using a pair of PWs in series with an output voltage accurate to within 12\%. PWs can also transmit analogue and digital data with a frequency of up to 19 kHz, which with the addition of a buffer, can drive high current circuits. We have demonstrated that PWs can last approximately two months, a 4  fold increase on previous literature. Protoplasmic tubes can be modified with the addition of conductive or magnetic nano-particles to provide changes in functionality. \\
\emph{Conclusion}
This work has documented novel macro-scale data transmission through biological material; it has advanced the field of bio-electronics by providing a cheap and easy to grow conducting bio-material which may be used in future hybrid electronic technology.
\end{abstract}

\section{Introduction}

The concept of integrating biology and electronics has existed for decades, with biosensors using biological transducing elements with electronic conditioning for measurement of an analyte since 1956~\cite{palchetti2010biosensor}. Today the field of biosensors is growing, with medical applications such as glucose monitoring in patients with diabetes \cite{Renaud2014} a prominent field for research. Researches are starting to investigate the possibility of using biological elements as components in electronics, computers and more recently molecular wires \cite{Deutscher2013,Magoga1997,Ratner1998,Tian1998,Ito2004}. 10 mm lettuce seedlings have been investigated, demonstrating a passive resistance of 2.76 MOhms when used in low current applications  \cite{Adamatzky2014}. The bacteria \emph{Geobacter sulfurreducens} produces conductive biofilms and nanowires which may be used in a variety of engineering applications \cite{Lovley2011,Reguera2006}, from nanoscale electronic circuits to increasing the current produced in biological fuel cells. \\
\\

The use of biological components in electronics could provide renewable and cheap passive devices with unique properties based on their inherent structure and metabolism. . The plasmodium of \emph{Physarum polycephalum} is a large single celled protistic organism of the class \emph{Myxomecetes}; it grows in the form of a network of protoplasmic tubes which extend several centimetres between sources of food, often found in dark damp wooded areas where it digests decomposing organic matter \cite{Stephenson2000}. This particular organism demonstrates a basic intelligence, where it forages for food while avoiding repellent stimuli \cite{Ueda1975,Whiting2014c} the organism optimises its network of protoplasmic tubes to connect all local food sources with minimum biomass \cite{Adamatzky2010}. In addition to chemotaxis, it displays phototaxis, thermotaxis and thigmotaxis \cite{Adamatzky2014b,Adamatzky2013,Hader1984,Mayne2014,Nakagaki1999,Whiting2014,Whiting2014a,Whiting2014b,Whiting2014c,Whiting2014d}. Using these behavioural paradigms, it is possible to route growth of protoplasmic tubes into certain networks and shapes \cite{DeLacyCostello2013}, such routing demonstrates the organism’s ability to solve mazes using the shortest path to find food sources of chemoattractants \cite{Nakagaki2000}. Several studies have used this shortest-path approximation to model worldwide transport networks on flat and three dimensional landscapes \cite{Adamatzky2010a,Adamatzky2012}, revealing a strong similarity between networks produced by organism and those currently in place.

In 2013 Adamatzky experimentally demonstrated~\cite{adamatzky2013physarum} that  protoplasmic networks can play a role of self-growing and self-repairing conductors, which can be insulated and, in principle, grown on conventional silicon hardware. In the present study we advance these results towards practical implementations and applications. Recent studies by Adamatzky \cite{Adamatzky2014a,adamatzky2014route} and Whiting \cite{Whiting2015e} have documented the use of protoplasmic tubes of \emph{Physarum polycephalum}, dubbed Physarum Wires (PW); the literature investigates the voltage conducting properties of PWs with a view to using them as a biotic component in electronics and computers. These papers demonstrated that 1cm PWs grown between agar hemispheres have a high resistance of approximately 2-3 MOhms. Initial work reported that these PWs could pass DC voltages with reasonable attenuation, proceeding to document that the PWs allowed the conduction of alternating frequencies with an attenuation profile similar to a low pass filter. However it is worth noting that the method of high frequency attenuation is not known; it was suggested that either the cytology of the organism was responsible or that there was some unavoidable parasitic capacitance inherent in the equipment set up. Adamatzky’s work demonstrated that the PWs could be insulated by coating the protoplasmic tube in silicone gel with little or no observable ill effects on the organism. Another very interesting aspect pointed out in these papers, is the ability of these PWs to repair themselves when physical trauma had severed the tube within a matter of hours, restoring the conductive pathway. Novel wires have been developed which are capable of reconnecting the conductive pathway if broken or severed \cite{Palleau2013}, however despite the authors' claims, they cannot self-heal, as they require the wire to be physically reconnected, after which they reform a metallic conductive pathway. Other self-repairing conductive pathways such as those reported by Williams et al. \cite{Williams2007} can repair micro-cracks in the structure at a molecular level however the ability to completely repair a fully severed gap is not reported. Another limit of these self-repairing wires is that they have to be encapsulated in some form of polymer which contains the self-repairing constituent; they are also not capable of bridging a gap, unlike PWs. Studies conducted by Mayne \emph{et al.} \cite{Mayne2014a,Mayne2013} have demonstrated that \emph{P. polycephalum} can also be cultivated in the presence of   a variety of different bio-compatible nano and micro particles with conductive and magnetic properties which are internalised via the organism’s inherent feeding mechanisms. This in effect functionalises consequent PWs grown; desirable properties such as enhanced conductivity and alterations in membrane potential dynamics can be bestowed on a functionalised PW where such properties are desired It is envisaged that loading of specific materials may provide a known change in resistance which may aid in the production of organism-based electrical components. It is these features which is one of the biggest advantages of using a PW, their self-repairing properties \cite{adamatzky2013physarum}, self-routing capability \cite{Adamatzky2010a,adamatzky2013physarum,Adamatzky2012,Nakagaki2000} and the ability to grow on a significant range of substrates \cite{Adamatzky2010}, means that PWs could be used for conductive pathways in conditions which require high connection reliability in hostile conditions as successfully as traditional electronics.

It is the purpose of this paper to document the first practical electronic circuits using Physarum Wires, along with identifying and addressing several limitations mentioned in previous literature. 

\section{Method}

\subsection{Cultivating \emph{Physarum Polycephalum}}

\emph{Physarum polycephalum} plasmodium was grown on non-nutrient 2\% agar gel in 9 cm diameter Petri dishes (Fisher Scientific, UK); the culture was fed with a small amount of microwave-sterilised organic rolled oat flakes (Waitrose, UK) every 2 days. The plasmodium grew accross the agar surface towards the oat flakes which it colonised and digested; after 1 week the plasmodium was transplanted to new agar by transferring a colonised oat flake, in order to limit unwanted microbial growth. The plasmodium colonised agar Petri-dishes were kept at room temperature in a dark cupboard, sealed with Parafilm Wrapping Film (Fisher Scientific, UK) to maximise moisture retention of both the agar and plasmodium. 

\begin{figure*}[!tbp]
\centering
\includegraphics[width=0.9\textwidth]{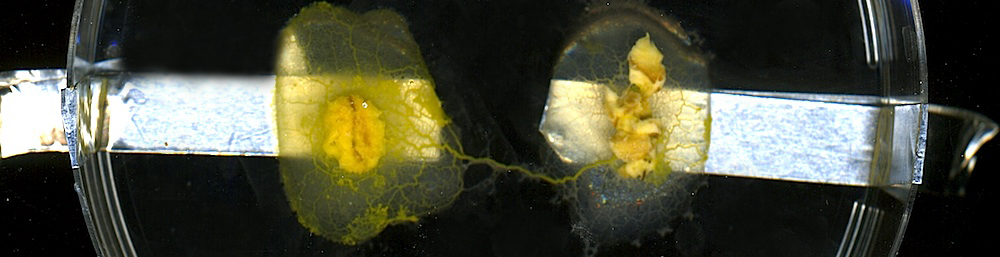}
\caption{An example Physarum Wire grown between agar hemispheres.}
\label{fig:tube}
\end{figure*}

Physarum Wires (PW), see example in 
Fig.~\ref{fig:tube} were produced on 9 cm Petri-dishes in accordance with the methods outlined previously \cite{adamatzky2013physarum,Whiting2014c}; a pair of 1ml 2\% agar hemispheres were placed on two strips of aluminium conductive tape (each were 10cm long and 1cm wide), at the centre of the petri-dish such that the closest edges are one centimetre apart (figure PhyWire), a P. polycephalum cultured oat flake was placed on one hemisphere and a non-colonised oat flake was placed on the other, a protoplasmic tube then grew between the agar hemispheres in most instances.

\subsection{Physarum Wire Lifespan}

The lifespan of the plasmodial stage of PP is limited by environmental conditions such as temperature and humidity, as well as food supply, contamination or microorganism competition. Even the presence of light can induce sporulation; reliability of sporulation is increased when certain chemical triggers are present \cite{Daniel1962}. Should the organism or supporting growth media dry out, the organism will enter a sclerotial stage where it dries out and becomes hard and immobile; this dormant stage can also be triggered by excess heat \cite{Whiting2014b} whether in a passive manner by plasmodia dehydration or an active self-preservation mechanism in the organism. Placing active plasmodium on certain substrates such as silica has demonstrated sclerotia formation , however it is unclear if this is due to lack of nutrition or a property of the surface such as wettability, material charge or surface roughness. While no investigative studies have been performed on surface properties and the growth of plasmodia, the authors suggest that this may have an effect on the growth and health of plasmodia. Should ideal conditions for all parameters be at the optimum level, then it is hypothesised that a culture of plasmodium could be maintained indefinitely, however it is very difficult to maintain optimum conditions. Continual feeding of the plasmodia requires a periodic addition of oat flakes, it is almost impossible to avoid the contamination of the plasmodia, air or growth medium in the culture vessel at this stage. Another limitation is the spatial memory of \emph{Physarum polycephalum}; when navigating a surface, the organism deposits extracellular slime which the searching front of pseudopodia avoids when searching for new food sources \cite{Reid2012}. The spatial deposition means that the area of viable growth medium decreases over time, which may limit the plasmodial life cycle. 

It has been reported in several papers that the life of a protoplasmic tube grown in this manner is between 3 and 7 days  \cite{adamatzky2013physarum}, limited by microbial growth on the agar and oat flake as well as drying of the agar hemispheres and plasmodium. The current maximum lifespan is not ideal for building computers and electronics from slime mould, so to extend the duration of operation several variables in production and storage of PWs were varied. It has been noticed by the authors that plasmodial cultures on agar grow more slowly when stored in colder conditions; colder temperatures also minimise the loss of moisture from the agar \cite{Smith2012}, therefore the temperature of fully grown PWs was varied. The authors have also noted that high humidity conditions tend to inhibit sclerotia formation, which was investigated.

\subsection{Using Physarum Wires With Buffers}

PWs have a high resistance of approximately 2 MOhms, therefore in practical electronics circuits they have limited use; high value resistors are used in such instances as ESD grounding, charge sensitive feedback devices, frequency filters, or other applications where low levels of current flow are essential. Medical devices which interface directly with a human subject require high resistance isolation to avoid potentially damaging or deadly currents passing to the patient in situations of device failure; the PWs pass waveforms of up to 19KHz with no distortion  \cite{Whiting2015e}, while most human electrical signals such as Electro-Encephalography (EEG) occur between 0 and 100 Hz. One problem of having such a high resistance when conducting signals is the high impedance of the signal output of the PWs; in order to have low impedance and to be able to conduct the input signal, a unity gain buffer may be used. The unity gain buffer has high input impedance and low output impedance, and also allows for very high current gains, enabling the driving of high current systems such as speakers (1W 8 Ohm 4cm, RS Components, UK), super-bright LEDs (5mm 12000 mcd Blue LED, RS Components, UK) or even motors (7W 4.5-15V 7300 rpm DC motor, RS Components, UK). Since the waveform is unchanged, the combination of a PW followed by a unity gain buffer can conduct high current waveforms like a traditional copper wire. 

\subsection{Using Physarum Wires To Transmit Digital And Analogue Data}

\begin{figure}[!tbp]
\centering
\includegraphics[width=0.49\textwidth]{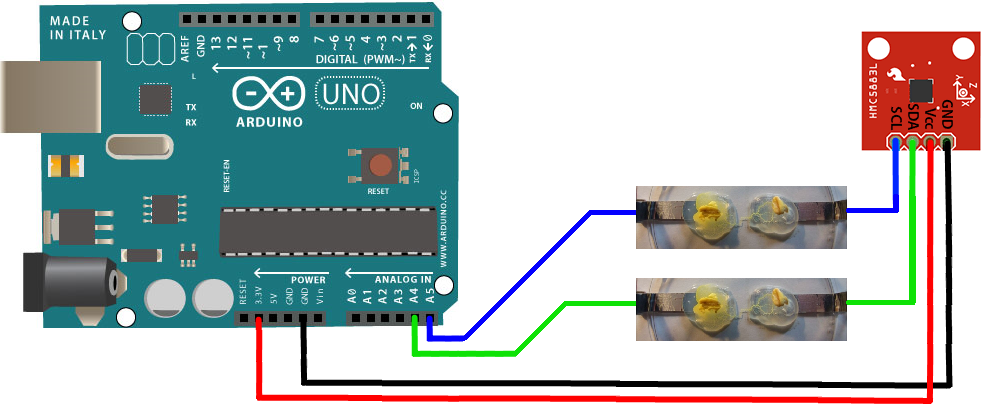}
\caption{Physarum Wires facilitating I2C communication between an Arduino Uno microprocessor and a 3-axis digital compass.}
\label{fig:PhysarumArduino}
\end{figure}

Digital systems are crucial to the modern world, with computers and the internet being a focal technology in society; the ability of a PW to pass a digital waveform is therefore important if it is to be integrated into future wetware computer and electronics systems. An important advantage of digital systems is their high noise rejection quality, which allows for error-free binary information transmission and processing in conditions where analogue signals would be very susceptible to noise; the authors tested PWs with traditional digital communication systems. An Arduino Mega (RS Components, UK) was programmed to communicate with a slave Digital 3-axis Compass (HMC5883L. Honeywell, NJ USA) using I2C protocol. The equipment was set up as shown in figure \ref{fig:PhysarumArduino}, with the Physarum Wire used as the digital bus for data communications on the SDA and SCL lines. Communication between the Arduino microprocessor and the magnometer was established and data was transferred between the master and slave devices. Valid magnometric data was confirmed by movement and rotation of the magnometer and subsequent change in data received along the PW bus. Baud rate was increased in stages from 1200 Bd until data was too corrupted to facilitate successful data transmission.

In addition to digital signals, analogue signals are important to many electronic systems, almost all data that we collect from the world is in analogue form, which gets converted into digital binary information for transmission and computation, it is important to maintain the integrity of the original signal in order to most accurately represent the original signal. PWs have already been shown to conduct analogue signals below 19 KHz with no distortion and may be used in the transmission of analogue signals; signals above the cut off frequency are attenuated as the frequency increases. Some analogue computers and circuits require the low-pass filter properties, therefore this property may be ideal for certain situations. The authors have transmitted audio signals along PWs to speakers with very accurate audio reproduction. 

\begin{figure}[!tbp]
\centering
\includegraphics[width=0.5\textwidth]{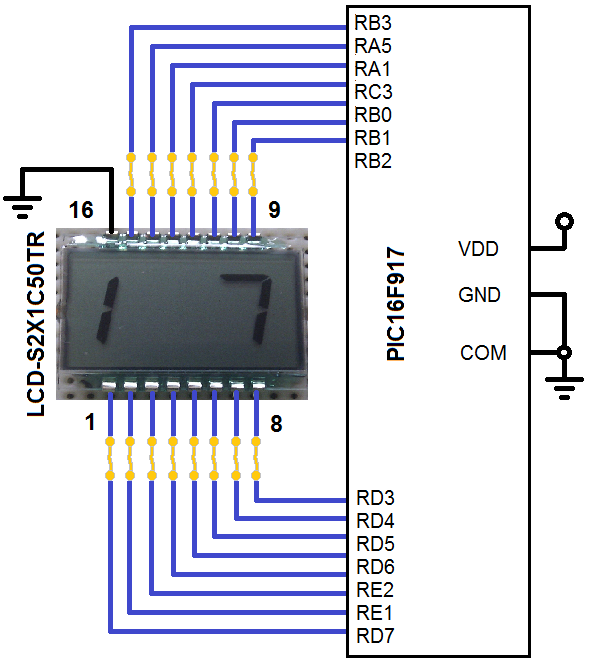}
\caption{A LCD screen illuminated by microprocessor control using an array of Physarum Wires}
\label{fig:PhysarumLCD}
\end{figure}

Due to the current restricting nature of the PWs documented in previous literature, the authors have investigated the possibility of using components which require very low levels of current. Alphanumeric LCD screens (Lumex LCD-S2X1C50TR, Farnell, UK) were illuminated by passing a DC voltage along an array of PWs to the display’s pins from a PIC16F917 microprocessor, as shown in figure \ref{fig:PhysarumLCD}. Low current LEDs (Avago Technologies HLMP-D155, Farnell, UK) were also tested with a 6 volt battery and a PW.


\subsection{Physarum Wires As Voltage Dividers}

\begin{figure}[!tbp]
\centering
\includegraphics[width=0.5\textwidth]{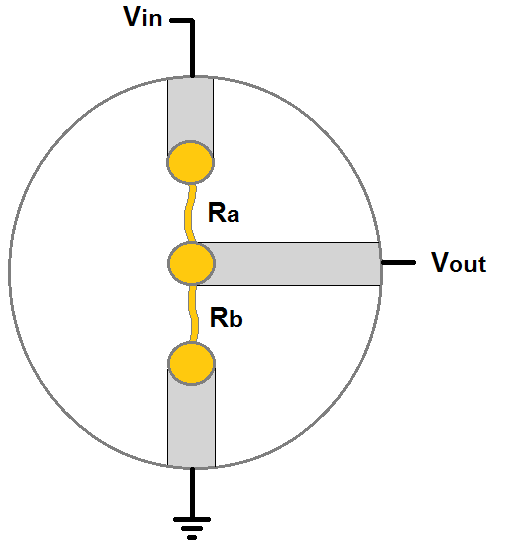}
\caption{A voltage divider created by two connected in-series Physarum Wires}
\label{fig:2wirevoltagedivider}
\end{figure}

An example of a practical circuit using resistors is the voltage divider; a passive linear circuit using two or more resistors (other passive components may be used but resistors are the most common), an output voltage which is lower than the input can be created. The voltage output, V$_{out}$, for a two resistor potential divider is proportional to the input voltage, V$_{out}$, and relative resistance values as shown in equation 1. They are used in many different analogue and digital systems for power supplies, reference voltages or circuit biasing; variable resistors are also used for variable control of the output voltage. Physarum voltage dividers were produced using two PWs grown between three agar hemispheres, as shown in figure \ref{fig:2wirevoltagedivider}; a variety of input voltages from 1 to 12 volts were applied using a Digimess Variable DC power supply (RS Components, UK) and the output voltage was measured and compared to the expected output of an ideal potential divider using a Isotech digital multimeter (RS Components, UK). 

Occasionally two protoplasmic tubes would grow between the same pair of agar hemispheres; the voltage divider set-up was still tested using this growth to determine the effect of multiple PWs between the same agar hemispheres. Scheme A was measured with R$_a$ as two protoplasmic tubes and scheme B was measured with R$_b$ as two protoplasmic tubes.

\begin{center}
\(V_{out} = {\frac{R_b}{R_a + {R_b}}} \times V_{out}\) \end{center} \begin{flushright} Equation 1 \end{flushright}

\begin{figure*}[!tbp]
\centering
\includegraphics[width=0.9\textwidth]{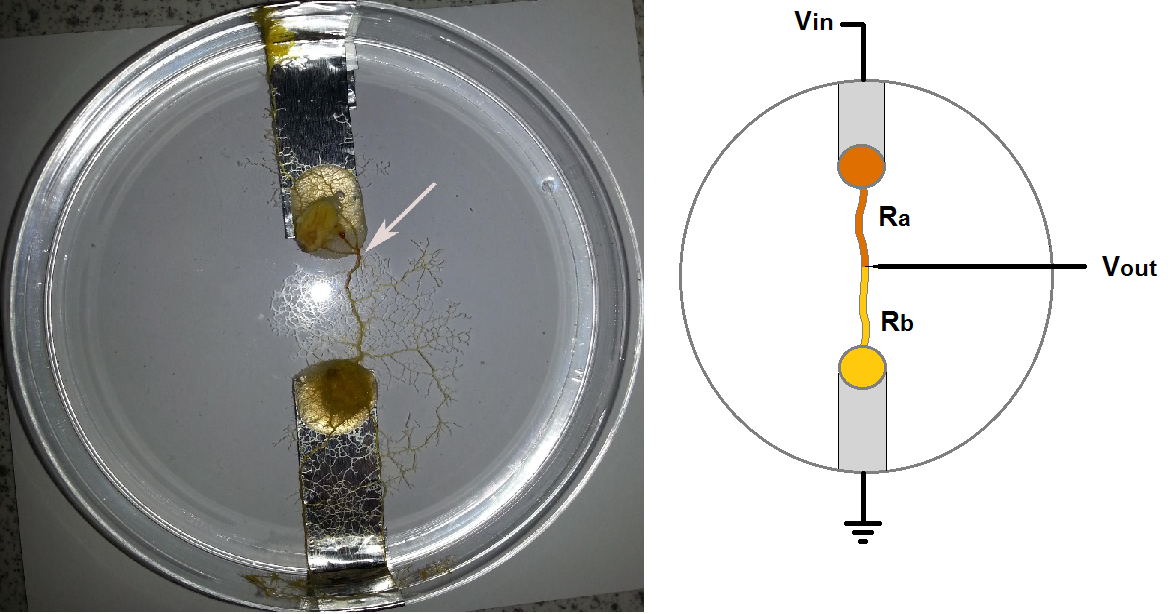}
\caption{(Left) Photograph to demonstrate functionalization of half of a PW by microinjection with biocompatible magnetite NPs. (Right) Schematic diagram to demonstrate how a voltage divider may be generated through the use of semi-functionalised PWs using a needle electrode.}
\label{fig:NpVoltageDivider}
\end{figure*}

A second variety of voltage divider utilising PWs may be produced by using plasmodial tubes functionalised with metallic nanoparticles (Fig. \ref{fig:NpVoltageDivider}). In laboratory investigations, it was found that by piercing a PW with a glass microinjection capillary filled with a solution of starch-coated 100nm magnetite (Iron II/III oxide) nanoparticles (Chemicell, Germany), a small quantity of solution would dispense into the tube which was subsequently transported up half of the PW in the direction of cytoplasmic flow. As this  is followed by a temporary cessation in shuttle streaming (as is usual for plasmodial vessel injury) it leaves a semi-functionalised wire with two different conductivity profiles. If the microelectrode were then used as a probe, this would create a voltage divider without the need to grow multiple tubes.

\section{Results}

\subsection{Increasing The Lifespan Of Physarum Wires}

Maintaining a Physarum Wire in a cold and humid environment increased the operational life from 2 weeks as previously reported in the literature, to 8 (currently) weeks. This extension was performed by managing and maintaining conditions of growth in the organism so to minimise drying of either agar medium or the PW itself; the optimal conditions for life extension are maintaining culture at temperatures between 6 and 12$^o$C while maintaining humidity and regular feeding to sustain a constant source of food for the organism. The organism was fed weekly, with 1 microwave sterilised oat flake placed on each agar hemisphere; at the same time, the inside of the Petri-dish lid dish was placed over steaming water, to enable condensation to form for humidity, the lid was then replaced and the Petri-dish was re-sealed with Parafilm. Interestingly, keeping the organism in dark or light conditions had no significant effect on the lifespan. It should be noted that oscillation does not always occur when the PW is maintained in these conditions, this could be due to the decreased temperature or effective increase in lifespan. The oldest tubes were tested for functionality and were still conductive enough to drive alphanumeric LCDs and Low current LEDs, as described below.

Cultures which did not have high humidity using the steam method yet were incubated at room temperature often succumbed to microbial growth on the agar hemispheres, lasting between 15 and 24 days, however cultures which were kept in cool conditions but without humidity tended to dry out despite the lack of warmth, and lasted 12-20 days. Cultures which were fed weekly lasted longer than the unfed counterparts in all combinations of temperature and humidity. 

\subsection{Using Physarum Wires With Buffers}

Using a PW as a passive element of high resistance is often limiting, adding a buffer provided the conduction of voltage waveforms with high output current, as the buffer acts as a high gain current amplifier with unity voltage gain. We have tested a number of applications using the PW and buffer and found that the circuit was sufficient to drive a super-bright LED in DC and pulsing modes, drive an 8ohm speaker and piezo buzzer with simple and complex audio signals offering audio fidelity very similar to the source quality. The standard human hearing range is between 20Hz and 20KHz, so within the pass filter limits of the PWs. Without the addition of the buffer, LED and speaker driving is significantly lower, with LEDs being very low brightness and speakers being very low volume. 

\subsection{Using Physarum Wires To Transmit Digital And Analogue Data}

Physarum Wires can pass binary signals between digital devices, as the authors demonstrate in this paper. An Arduino Mega (RS Components, UK) was programmed to communicate with a slave Digital 3-axis magnometer using I2C protocol. The equipment was set up as shown in figure \ref{fig:PhysarumArduino}, with the Physarum Wire used as the digital bus for data communications on the SDA line. Communication between the Arduino microprocessor and the magnometer was established and data was transferred between the master and slave devices. Valid magnometric data was confirmed by movement and rotation of the magnometer and subsequent change in data received along the PW bus. Baud rates of 1200, 2400, 9600, 19200, 38400 and 115200 were tested, with no reported corruption or invalid data transmission at baud rates of 19200 and lower, invalid data was received sporadically at 34800 baud rate, however mostly valid and confirmed data was received; using the baud rate of 115200 most of the data was invalid and the data was unusable.

This paper has reported that analogue voltage signals can be passed along a Physarum Wire; at frequencies below 20kHz analogue signals pass through the PW with minor attenuation. It has been established by practical experiments with electronic circuits that signals pass through the PW as though it were a passive resistor in the pass-band range; this was demonstrated by playing audio signals though the PW to a speaker and Piezo buzzer where the signal was transduced into an audio signal. As frequency increases, attenuation increases, above the cut off frequency; analogue signals are less useable above the cut off frequency. It has been recently demonstrated that \emph{P. polycephalum} can be loaded with conductive micro- and nanoparticles \cite{Dimonte2014,Mayne2013,Mayne2014a} which would increase the efficiency of conduction by decreasing resistance. Filling protoplasmic tubes with defined volumes of conductive material has the potential to create known value discrete components such as resistors and inductors. It has also been suggested that PWs can be used in biological analogue computers \cite{adamatzky2013physarum,Whiting2015e} as they integrate analogue signals.

Physarum Wires can be used to illuminate low current LEDs with a 9V DC voltage source. Larger source voltages can increase the illumination by increasing the current through the PW in accordance with Ohm’s Law, as the increase in voltage drives an increase in current through the PW. PWs can also be used to illuminate reflective LCD alphanumeric display screens; the LCD-S2X1C50TR module was illuminated with an array of PWs connecting each pin of a PIC16F917 microprocessor which controlled the illumination of each individual segment of the LCD screen. Each LCD segment was tested and showed that even with a small amount of current passing through the PWs, the segments were illuminated.

\subsection{Physarum Wires As Voltage Dividers}

Potential dividers were produced using two PWs, which produced results which would be expected of an electronic resistor potential divider as described by equation 1. Assuming both PWs had equal resistance, the output voltage should be half the input voltage; this was the case using PW based potential dividers as shown in figure \ref{fig:2wirevoltagedivider}. The mean output is 12\% (SD 2\%) less than the expected output, however it is linear, suggesting that there is a marginal difference in resistance between the two tubes.

\begin{figure}[!tbp]
\centering
\includegraphics[width=0.49\textwidth]{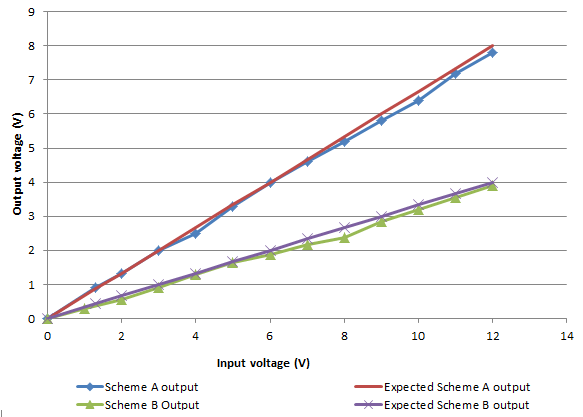}
\caption{Physarum Wire potential dividers with unequal resistances.}
\label{fig:schemeAB}
\end{figure}

When testing the potential divider as described by scheme B, the output appears to correlate with that expected of a voltage divider where the resistance of Rb is half that of Ra; this is plausible because two PWs in parallel would follow Kirchoffs law, leading to a halving over total resistance. Due to the novel nature of this particular growth, this set-up was tested both in the manner described by scheme B, but also with the petri-dish reversed so the two parallel PWs were in place of Ra, scheme A, to test the bi-directionality of such a growth. Figure \ref{fig:schemeAB} shows the output of these tests.


\section{Discussion}

\subsection{Physarum Wire Lifespan}

The authors have demonstrated the functional increase in PW lifespan from 2 to 8 weeks; this 4 fold increase in lasting functionality minimising one of the initial limitations of PWs. The authors have shown that environmental conditions are a limiting factor to PW lifespan and by decreasing temperature while increasing humidity, a dramatic increase in working life has been achieved. When maintaining low temperature conditions, the main problem for the lifespan of PWs changed from agar or organism drying, to one of microbial grown and contamination; production and weekly feeding of the Petri-dishes may allow unwanted microbe spores to enter the set-up, which is undesirable as this leads to competition with \emph{P. polycephalum} and ultimately withdrawal or sclerotia formation of the protoplasmic tube. It is suggested that PW Petri dish production in a clean room with microbial controlled environment would stop microbes contaminating the agar hemispheres, leading to increase in PW lifespan. Since opening the Petri-dish for weekly feeding may allow microbes to enter the culture, a slow release food source or cleaner method for introducing new sources of food would be desirable and lead to further yields in PW lifespan.
Low temperature conditions for PWs is not ideal, however given the specific benefits of significantly extending the lifespan of a protoplasmic tube, it is desirable to maintain a cool temperature while in use. Maintaining a constant low temperature is not uncommon in computing and electronics, however it would be desirable if passive cooling methods were employed to maintain a lower temperature, eradicating the need for a cooling power source.

\subsection{Physarum Wire length}

It has been noted in many papers \cite{adamatzky2013physarum,Whiting2014a,Whiting2014c} that typical \emph{P. polycephalum} protoplasmic tubes grown on plastic ranges from 5mm to 10mm. Macroscopic electronics circuit boards often require copper tracks which are longer than this, so it would be ideal to know the maximum length of PWs when designing hybrid circuit boards. It is known that \emph{P. polycephalum} can grow to significant lengths when on agar medium, however the Agar gel would interfere with the conduction properties of the protoplasmic tubes, therefore the authors document only long tubes grown on insulator material.  

In \cite{Mayne2014}, however, it was found that optically-coupled slime mould logical gates based on the previously described experimental environment for producing and measuring the electrical properties of PWs, both fail to propagate more frequently and take longer to propagate when the distance between electrodes is increased. This indicates that \emph{P. polycephalum} finds travelling across expanses of plastic unfavourable --- likely due to the relatively hydrophobic surface and lack of moisture or food available in/on the substrate, although variations in environmental humidity and the presence of other repellent factors in the experiment are likely to accentuate these issues --- and hence that to produce longer PWs with an appreciably high success of propagation rate, modifications to the experimental environment are required to make it more hospitable.

A PW is not limited to the length of 10mm and may easily be produced up to 10x longer experimentally: as demonstrated in our previous works on imitating road networks, the slime mould confidently propagates along bare plastic and navigates around elevations  on 3D printed nylon templates~\cite{adamatzky2014route}; the longest documented recorded protoplasmic tube exceeded 15 centimetres. These longer tubes often have multiple branches, as was desired in the previous work; branching may be desirable in order to create circuit nodes or nets. Growth and branching of the PWs will have to be closely controlled, as multiple parallel or branched tubes would change the functionality of the circuit. These increases in length show that practical circuit boards of significant size can be developed; this would be a significant advancement for the field of bioelectronics and hybrid wetware-hardware electronics.

Protoplasmic tube morphology is determined by environmental factors \cite{Takamatsu2009}, which is of importance when producing circuit boards using \emph{P. polycephalum}; while increasing the practical length of connections is important, the circuit layout could be optimised to facilitate the least biomass for any given circuit. We already know that \emph{P. polycephalum} uses the shortest path between multiple food sources, so routing of a circuit board could be optimised with attractant and repellent materials in accordance with Costello et al \cite{DeLacyCostello2013}. Tubes which are close or converging have a tendency to merge, which may cause problems for circuit boards, however if controlled, would be useful for producing circuit boards with track nodes or nets.

\subsection{Using Physarum Wires to transmit Digital and Analogue Data}

It has been shown that Physarum Wires can be used to transmit digital data at baud rates of 19200 or less; baud rates of 38400 showed some corruption of digital data, although most data packets were transmitted through the PW with no ill-effect. Tested baud rates above 38400 were very poor at transmitting digital I2C data. The transmission of digital data through the PWs shows that they may be used in digital systems using low frequency digital data.
Analogue data below 19kHz was also transmitted through PWs with very little attenuation; frequencies above 19kHz were increasingly attenuated as frequency increased. The standard human hearing range in normal conditions is between 20Hz and 20kHz which lies perfectly within this range, meaning Physarum Wires could be used in audio signal transmission without important signal loss; PWs could also be used as a filter in between a microphone and analogue-to-digital converter as it would passively filter out unwanted analogue noise outside the human hearing range.
PWs can also be used to transmit analogue and digital signals to illuminate low current LEDs and LCDs as demonstrated in this article; previously they had illuminated higher current LEDs \cite{adamatzky2013physarum}. The large resistance of the PWs does limit the current passing through them, however using unity gain buffers after the PWs, the current may be increased to usable levels as demonstrated by powering speakers or bright LEDs.
Current limiting devices are also important in medical electronics to protect the patient and operator from potentially harmful current levels; PWs have an inherent current limiting property and so could be used for this application.
The method of conduction in Physarum Wires is unknown, however it is suggested to be a property of electrolytic conduction using ions in the plasmodial fluid \cite{Whiting2015e}; the frequency limiting portion is believed to be the result of the cytoskeleton limiting speed of flow of the free ions in electrolytic solution.

\subsection{Physarum Wires as Voltage Dividers}

We have shown that two separate PWs can be used to create a voltage divider with an accuracy of 12 \% or better. The error is linear and is most likely due to differences in tube resistance; it was assumed that two tubes have equal resistance while calculating the ideal voltage, however resistances of PWs vary by 23\% RSD \cite{Adamatzky2014a}. Two PWs in parallel equates to two traditional resistors in parallel, giving half the resistance, which is an important finding in bio-electronics and computing, as biologically derived electronic components act like traditional electronics.

As has been documented by Mayne \cite{Mayne2014a,Mayne2013}, \emph{P. polycephalum} can absorb conductive nanoparticles via endocytosis; it has been shown that these conducting particles can change the resistance of the tubes, which may facilitate the production of a variety of dose-dependant resistances for fixed or variable resistors. Moving the position of the needle electrode would change the ratio of the resistors, much like a variable resistor, facilitating the rapid change of resistance and voltage output. Previous work \cite{Mayne2014a} involved loading protoplasmic tubes  with 100nm starch coated magnetite decreasing their resistance; the lowest value of resistance produced using this method was 18 kilo Ohms, which is within the region of commonly used resistors.

This work has documented the production of voltage dividers using either two series PWs, or one PW with a needle electrode along its length. This has demonstrated that real electronic circuits can be produced using PWs and has the potential to augment an already growing field of electronics derived from biology \cite{Renugopalakrishnan2006}.

\bibliographystyle{spphys}
\bibliography{references}
\end{document}